\documentclass[12pt]{article}

\usepackage[utf8]{inputenc}
\usepackage{charter}
\usepackage{fullpage}
\usepackage{hyperref}
\usepackage{graphicx}
\usepackage{lipsum}
\usepackage[sort&compress,numbers]{natbib}
\usepackage{hyperref}
\hypersetup{hidelinks}
\usepackage[total={6.5in,9in},top=1in,headsep=0.1in,headheight=1in]{geometry}

\usepackage{color}
\usepackage{amsmath,amssymb,amstext}
\usepackage{latexsym}
\usepackage{aas_macros}

\usepackage{xspace}

\newcommand{\LCDM}{$\Lambda$CDM\xspace}

\def\ltsima{$\; \buildrel < \over \sim \;$}
\def\lsim{\lower.5ex\hbox{\ltsima}}
\def\gtsima{$\; \buildrel > \over \sim \;$}
\def\gsim{\lower.5ex\hbox{\gtsima}}

               {\refstepcounter{theorem}\vspace{2ex}%
                \noindent{\bf Examples \thetheorem} }%
               {\hfill $\diamond$}

               {\refstepcounter{theorem}\vspace{2ex}%
                \noindent{\bf Example \thetheorem} }%
               {\hfill $\diamond$}

\newcounter{problem} 
\renewcommand{\theproblem} {\arabic{problem}}
               {\refstepcounter{problem} \vspace{2ex}%
                \noindent{\bf Problem \theproblem} }%
               { }

               {\noindent{\bf Solution}}%
               {\hfill $\Box$ \\[1ex]}

\newlength{\boxwidth}

\newlength{\fullboxwidth}
\newlength{\fullinboxwidth}
\newcommand\snowmass{
\begin{center}
  \rule[-0.2in]{\hsize}{0.01in}\\
  \rule{\hsize}{0.01in}\\
  \vskip 0.1in
  Submitted to the Proceedings of the US Community Study\\ 
  on the Future of Particle Physics (Snowmass 2021)\\
  \rule{\hsize}{0.01in}\\
  \rule[+0.2in]{\hsize}{0.01in}\\[-2em]
\end{center}
}

\usepackage[firstpage=true]{background}
\backgroundsetup{contents={\parbox{6.5in}{\snowmass}}, scale=1,placement=top,opacity=1,color=black,position={3.25in,1.2in}}

\usepackage{fancyhdr}
\fancypagestyle{plain}{%
  \fancyhf{}%
  \fancyhead[C]{}
  \fancyfoot[C]{\thepage}
}

\fancypagestyle{empty}{%
  \fancyhf{}%
  \fancyhead[C]{{\it Cosmological Simulations and Modeling}}
  \fancyfoot[C]{\thepage}
}
\title{Snowmass2021 Computational Frontier White Paper: Cosmological Simulations and Modeling}
\date{}

\usepackage{authblk}

\author[1]{Marcelo A.~Alvarez}
\author[2,3]{Arka Banerjee}
\author[4]{Simon Birrer}
\author[5,$\dagger$]{Salman Habib}
\author[5]{Katrin Heitmann}
\author[1,$\dagger$]{Zarija Luki\'c}
\author[6]{Julian B.~Mu\~noz}
\author[4]{Yuuki Omori}
\author[1]{Hyunbae Park}
\author[7]{Annika H. G. Peter}
\author[1]{Jean Sexton}
\author[8]{Yi-Ming Zhong}

\affil[1]{Lawrence Berkeley National Laboratory, Berkeley, CA 94720, USA}
\affil[2]{Department of Physics, Indian Institute of Science Education and Research,
Homi Bhabha Road, Pashan, Pune 411008, India}
\affil[3]{Fermi National Accelerator Laboratory, Batavia, IL 60510, USA}
\affil[4]{SLAC, 2575 Sand Hill Road, Menlo Park, CA 94025, USA}
\affil[5]{Argonne National Laboratory, Lemont, IL 60439, USA}
\affil[6]{Center for Astrophysics \textbar{} Harvard \& Smithsonian, 60 Garden Street, Cambridge MA 02138, USA}
\affil[7]{CCAPP, Department of Physics, Department of Astronomy, The Ohio State University, Columbus, OH 43210, USA}
\affil[8]{Kavli Institute for Cosmological Physics, University of Chicago, Chicago, IL 60637, USA}
\affil[$\dagger$]{Paper Facilitator}

\begin{document}

\maketitle

\begin{abstract}

Powerful new observational facilities will come online over the next decade, enabling a number of discovery opportunities in the ``Cosmic Frontier", which targets understanding of the physics of the early universe, dark matter and dark energy, and cosmological probes of fundamental physics, such as neutrino masses and modifications of Einstein gravity. Synergies between different experiments will be leveraged to present new classes of cosmic probes as well as to minimize systematic biases present in individual surveys. Success of this observational program requires actively pairing it with a well-matched state-of-the-art simulation and modeling effort. 
Next-generation cosmological modeling will increasingly focus on physically rich simulations able to model outputs of sky surveys spanning multiple wavebands. These simulations will have unprecedented resolution, volume coverage, and must deliver guaranteed high-fidelity results for individual surveys as well as for the cross-correlations across different surveys. The needed advances are as follows:
\begin{itemize}
\item Development of scientifically rich and broadly-scoped  simulations, which capture the relevant physics and correlations between probes  
\item Accurate translation of simulation results into realistic image or spectral data to be directly compared with observations
\item Improved emulators and/or data-driven methods serving as surrogates for expensive simulations, constructed from a finite set of full-physics simulations
\item Detailed and transparent verification and validation programs for both simulations and analysis tools
\end{itemize}
Stringent accuracy requirements will be imposed given the statistical power of the new datasets. Use of exascale and post-exascale computing resources, AI/ML methods, and robust error control mechanisms will be essential features of these efforts. This white paper details a simulation program to fully realize the potential of the coming decade’s cosmic observatories.

\end{abstract}

\section{Introduction}

The next generation of cosmology experiments~\cite{cmbs4,desi1,desi2,euclid,lsst,so,spherex} are aimed at exploring some of the most exciting questions in fundamental science -- the twin mysteries of dark energy and dark matter and the origin of primordial fluctuations, along with the use of cosmology as a probe of particle physics (e.g., studies of the neutrino sector or the nature of dark matter). Interpreting the results of many of these experiments, spanning measurements across multiple temporal epochs and length scales, involves solving an inverse problem, where given the observational results one wishes to unearth the details of the underlying physics. Modeling the effects of changes in parameter values as well as in the physical assumptions and establishing a direct connection to the observations across multiple surveys is a complex and challenging task. Cosmological simulations are the only way to approach this problem, simultaneously addressing the myriad issues associated with dynamical complexity, cross-correlations,  and strict requirements on error control.

The required ability to create simulated ``virtual universes'' on demand is the fundamental computational challenge faced by the Cosmic Frontier. Indeed, it is not an exaggeration to say that the ultimate scientific success of the next generation of sky surveys hinges critically on the success of the underlying modeling and simulation effort. 

The generation of these virtual universes can be accomplished in different ways \cite{somerville,Vogelsberger:2019ynw}. Large gravity-only simulations~\cite{angulo_hahn} are used as the backbone for building sky maps that closely resemble the observations from large surveys~\cite{cosmodc2,dc2}. This approach requires careful modeling to establish the ``galaxy-halo'' connection~\cite{Wechsler2018}. The modeling strategies range from simple methods that take limited information into account and rely on empirical modeling assumptions to elaborate schemes that try to model galaxy formation processes as closely as possible but without directly modeling computationally expensive gas physics and feedback effects~\cite{somerville}. Hydrodynamics simulations attempt to model galaxy formation in cosmological volumes including gas physics and feedback effects~\cite{Vogelsberger:2019ynw}. They employ phenomenological subgrid models whenever the dynamical range needed to resolve the physics of interest is too vast to start from first principles. The ultimate aim is to advance these different methods such that they all converge to the same answer -- faithfully describing our Universe in all observable wavebands. 

With the advent of exascale computing resources~\cite{ecp}, several opportunities will arrive, but taking full advantage of them will not be straightforward. The high-performance computing (HPC) system architectures, associated software ecosystem, and data infrastructure will be substantially different from that of the previous generation. Adjusting to this computational environment, along with its variety and rapid evolution, will require special attention and substantial human resources.  The resolution and volume of gravity-only simulations will enable the creation of ever more detailed synthetic sky catalogs. Hydrodynamics simulations in large cosmological volumes with a rich set of well-tuned subgrid models will be feasible. These simulations will allow us to study and mitigate possible systematic effects that might obscure fundamental physics insights. Synthetic skies will be developed across multiple wavebands and surveys (e.g., Ref.~\cite{dc2}). In order to realize this vision, we have to fully exploit the next generation of HPC resources for these large-scale simulations, and develop efficient analysis approaches to connect the simulations closely to observational data. The large data sets may require additional dedicated data-intensive computing resources to run complicated analysis workflows (potentially including cloud access).

\section{Numerical Simulations}

Numerical simulations play a critical role in delivering Cosmic Frontier science, both as the means to formulate precise theoretical predictions for different cosmological and astrophysical models, but also in evaluating and interpreting the capabilities of current and planned experiments. For optical surveys, the chain begins with a large cosmological simulation into which galaxies and quasars (along with their individual properties) are placed using semi-analytic or halo-based models. A synthetic sky is then created by adding realistic object images and colors and by including the local solar and galactic environment. Propagation of this sky ``image" through the atmosphere, the telescope optics, detector electronics, and the data management and analysis systems constitutes an end-to-end simulation of the survey. A sufficiently detailed simulation of this type can serve a large number of purposes such as identifying possible sources of systematic errors and investigating strategies for correcting them and for optimizing survey design (in area, depth, and cadence). The effects of systematic errors on the analysis of the data can also be investigated; given the very low level of statistical errors in current and next-generation precision cosmology experiments, and the precision with which deviations from $\Lambda$CDM are to be measured, this is an absolutely essential task.

\begin{itemize}
\item {\bf N-body simulations.}
Gravity is the dominant force on large scales, and dark matter outweighs baryons by roughly a factor of five to one.  Thus N-body simulations accurately describe matter fluctuations from the largest observable scales down to scales deep into the nonlinear regime.  Due to their computational efficiency, conventional N-body simulations (i.e., those treating cold dark matter models and some variants thereof) cover a wide dynamic range (Gpc to kpc, allowing coverage of survey-size volumes), with relative ease. It should be noted, however, that multi-Gpc-scale simulations at high mass resolution are still significantly expensive, even on exascale resources.

N-body simulations have essentially no free parameters, and when properly designed, can reach sub-percent accuracy over a wide dynamic range. A significant part of our current knowledge of nonlinear structure formation has been a direct byproduct of advances in N-body techniques. In the near future, survey-scale simulation suites are likely to be dominated by N-body simulations, although some large-volume hydrodynamic simulations will begin to appear at a reasonable mass resolution for the baryonic component.

The key shortcoming of the N-body approach is that the physics of the baryonic sector is not accounted for, thus many of the directly observable quantities are derived in somewhat heuristic ways and by adding a number of modeling or nuisance parameters.  Galaxies in N-body simulations are usually reconstructed by applying additional modeling on top of a simulation, such as the halo occupation distribution (HOD) \cite{Berlind2002}, sub-halo abundance matching (SHAM) \cite{Kravtsov2004, Hearin2013}, or semi-analytic modeling (SAM) schemes (for a description of many SAM approaches, see for example Ref.~\cite{Knebe2015}).

\item {\bf Hydrodynamical Simulations.}
The primary role of hydrodynamical simulations in cosmology is to provide a reasonably accurate description of the distribution of baryons, to quantify the effects of baryons on various probes of large-scale structure (e.g., galaxy clustering, weak and strong lensing, matter-galaxy cross-correlations, redshift-space distortions, Lyman-$\alpha$ forest, SZ signal, 21cm and other line intensity mapping signals), and to provide useful results for the distribution and properties of galaxies, groups, and clusters. Because the final results depend strongly on the choices made for parameterized subgrid models, there is substantial variability in the robustness of the results, depending on the nature of the cosmic probe under consideration. There is, therefore, considerable interest in melding the results of hydrodynamic simulations with empirical modeling of galaxy properties, in order to produce a set of predictive forward models that can plausibly cover a wide range of physical galaxy formation scenarios. These phenomenological models parameterize baryonic effects in a form that can be used directly in constraining the dark sector.

The exascale systems that will be available shortly -- and in the second half of the decade, post-exascale computing resources -- will allow hydrodynamical simulations to become significantly more useful in cosmology (as compared to qualitative interpretation of astronomical observations). In particular, it is expected that there will be a coming together of very small scale, high resolution simulations that currently aim to study the details of galaxy formation at the level of individual objects with simulations that aim to model billions of galaxies. The hope is that this confluence of methods, combined with new observations, will significantly improve the robustness of the obtained results.

\item {\bf Beyond $\Lambda$CDM Simulations.} Although $\Lambda$CDM has been very successful on large scales, the fact that dark energy is not theoretically understood and that at small scales different dark matter models may have different signatures that will be observationally accessible has motivated the development of simulations in different directions. Modified gravity simulations typically involve the solution of a nonlinear variant of the Poisson equation and are therefore significantly more expensive than traditional (N-body) Vlasov-Poisson solvers. Different dark matter models may require the addition of new treatments of local interactions, or may not be accessible to an N-body approach at all (as in the case of fuzzy dark matter models). For further details on the last topic, we refer the reader to a related White Paper on simulations focusing on dark matter~\cite{DMsimsWP}.

\item{\bf Radiative Transfer Simulations.}
Radiative transfer is playing an increasingly important role in astrophysics and cosmology. It is especially important for modeling reionization, which is thought to have occurred when the earliest generations of galaxies created photo-ionized bubbles that grew and merged until overlapping completely.  In addition to driving reionization itself, the ionizing radiation from these galaxies also affected their own evolution, as well as the density structure of the intergalactic medium between them. This resulted in a complex feedback loop in which small-scale effects were tightly coupled to radiation originating from a multitude of galaxies over vast cosmological volumes, e.g., Refs.~\cite{2015MNRAS.449.4380R,2022arXiv220205869L}. The demands on the dynamic range and accuracy of radiative transfer simulations will increase dramatically as observations reach further into the epoch of reionization. 

\end{itemize}

\subsection{Target Probes and Observables}

While simulations geared towards a particular probe satisfying the requirements of a specific survey have well-defined road maps, simulations capable of describing more than one probe, especially those consisting of more than one experiment, are less developed and discussed in the community. Since an immense amount of cosmological and astrophysical information could be extracted from combinations of observables from different surveys, the development of such simulations is of great interest.

\begin{itemize}
    \item {\bf Galaxy clustering/lensing, cluster clustering/lensing/counts.} These are the key observables for photometric galaxy surveys (see \cite{LSST2012, Euclid2011, WFIRST2019}). While the 2-point correlation function has been commonly used to measure these observables \cite{DES2018}, there has been an increased interest in applying higher-point correlation functions \cite{McBride2011} as well as alternate summary statistics that capture higher-order information \cite{Pratten2012, Liu2016, Banerjee2021} to extract additional signal from non-Gaussian density fields. Suites of simulations are needed to make predictions for these summary statistics since equivalent analytical frameworks do not exist. Additionally, multi-wavelength simulations of galaxies will allow us to test our detection/deblending pipelines, improve photometric redshift error estimations and validate shape measurements through cross-correlations \cite{dc2, Rhodes2017}.
    
    \item {\bf Spectroscopic galaxies.}\footnote{While physically there are no differences between photometric and spectroscopic galaxies, we separate these here since their implementation in simulations are significantly different.} Spectroscopic instruments \cite{Takada2014, Schlegel2019, Bundy2019, MSE2019, Ellis2019} measure redshifts, radial velocities, gas dynamics and chemical compositions of galaxies. Cosmological information will be extracted through Baryonic Acoustic Oscillation (BAO) and redshift space distortions (RSD) measurements \cite{GilMarin2020}. These galaxies are ideal for galaxy–galaxy lensing analyses \cite{Heymans2021}, as well as for calibrating photometric redshifts using the clustering redshift technique \cite{Davis2017, VanDenBusch2020}. When correlated with CMB temperature maps, the distribution of gas in low mass systems can be mapped out by using the kinematic Sunyaev Zel’dovich (kSZ) effect \cite{Schaan2016, Hill2016, Smith2018}. Additionally, the Lyman-$\alpha$ forest can be used to measure the three-dimensional power spectrum to intermediate redshifts \cite{Bautista2017}, which can be correlated with galaxy/CMB lensing \cite{Doux2016}.

    \item {\bf CMB Lensing.} Lensing of the cosmic microwave background (CMB) measures the integrated mass between the last scattering surface and us. Experiments such as CMB-S4 will produce clean (i.e. polarization based) maps of the integrated mass at high detection significance \cite{Abazajian2016}. Since the signal is sensitive to the full redshift range of the observable Universe, it is correlated with all of the other probes listed \cite{Omori2019, Omori2019b}. It is especially useful for weighing distant objects that are beyond the redshift ranges accessible through optical weak lensing \cite{Geach2019}.
    
    \item {\bf Lyman-$\alpha$ forest.} Experiments such as DESI \cite{desi1} will observe the Lyman-$\alpha$ forest in the spectra of distant quasars $2 \lesssim z \lesssim 4$.  Statistical properties of the Lyman-$\alpha$ forest can be used to constrain thermal properties of the intergalactic medium \cite{Walther2019} and cosmological parameters \cite{NPD2015, Bautista2017}. The Lyman-$\alpha$ signal originates in low density regions, thus probing different parts of the universe from most other probes.

    \item {\bf tSZ/kSZ Effects.} Both the thermal (tSZ) and kinematic Sunyaev Zel’dovich (kSZ) effects are sensitive to the distribution of gas in the Universe. The SZ effects are strongly correlated with the locations of high gas densities such as in galaxy clusters, and are hence correlated with lensing \cite{Osato2020} and X-ray \cite{Hurier2014} observations. The kSZ signal can effectively probe the early universe as it also correlates with the ionization pattern of the intergalactic gas during reionization \cite{Park2013}.

    \item {\bf CIB.} The cosmic infrared background (CIB) consists of emission from dusty star forming galaxies at $z \sim 2$. The CIB is highly correlated with CMB lensing since their redshift kernels overlap well, and therefore the CIB has been used to delens the CMB \cite{Larsen2016, Carron2017}. The number counts and clustering measurements of these infrared galaxies as well as their properties such as stellar mass, star formation rate, dust mass, and metallicity can give us insights into galaxy evolution \cite{Maniyar2018, Simpson2020}, and are strongly related to the characterization of galaxies at lower redshifts \cite{Behroozi2013}.

    \item {\bf X-ray maps.} Experiments such as eRosita \cite{Merloni2012} will measure about $\mathcal{O}(10^5)$ clusters of galaxies and 3 million active galactic nuclei over the full sky. By exploiting the tight correlation between X-ray emission and mass, X-ray observations could be used to calibrate mass estimates of SZ-selected clusters \cite{Bulbul2019}.
    
    \item {\bf Line intensity mapping.} Experiments such as SPHEREx \cite{spherex} and SKA \cite{SKA2020} will map out the density field at $0.5 \lesssim z \lesssim 3$. While the treatment of foregrounds are anticipated to be challenging, density fluctuations of the dark ages could be measured cleanly by cross-correlating with CMB lensing maps. \cite{Tanaka2019}.
    
    \item{{\bf High-redshift 21-cm.}}
    Interferometers such as HERA~\cite{DeBoer:2016tnn}, or the SKA~\cite{Mellema:2012ht}, aim to give us access to 3D maps of the universe during cosmic dawn and reionization ($z \approx 5-30$), by using the 21-cm line of hydrogen. These maps track the density of hydrogen, processed by a factor that depends on its spin temperature and ionized fraction~\cite{Furlanetto:2006jb,Pritchard:2011xb}. As such, they provide invaluable information on the thermal and ionization state of the IGM at high redshifts, which can be used to learn about dark matter, as it can cool the gas~\cite{Munoz:2018pzp}, heat/ionize it~\cite{Lopez-Honorez:2016sur,Liu:2018uzy}, or delay structure formation~\cite{Sitwell:2013fpa,Munoz:2019hjh}.
\end{itemize}

\subsection{Modeling Challenges}
Cosmological simulations have a well-established history, going back to about half a century, when computers first became powerful enough to enable very early studies of structure formation in the universe~\cite{peebles}. Since then, progress has been rapid, and cosmological simulations now rank among the most complex and computationally challenging problems for HPC systems. This situation is likely to remain unchanged for the foreseeable future. Below we describe some of the modeling challenges that are being faced in the area of cosmological simulations. This is by no means a complete list, but it is generally representative of the type of advances that are needed.

\begin{itemize}
\item {\bf Volume/Resolution/Number of simulations.} A challenging aspect in generating simulations that encompass multiple probes, is the computational cost, as the base simulation needs to meet the precision requirements of all the individual observables.
In addition, some of the observables (such as tSZ/kSZ or Lyman-$\alpha$) require hydrodynamical simulations, which are computationally demanding. In estimating covariance matrices where a large number of realizations are essential, approximate methods or machine learning techniques \cite{Troster2019} to accelerate the simulation procedure will be required. In specific cases, the modeling of certain observables confronts a very large dynamical range. For instance, the 21-cm signal depends on X-ray and UV photons with long mean free paths ($\sim$ Gpc), whereas the first galaxies formed in very small haloes (with $M_h\sim 10^6 M_\odot$). In these examples, detailed hydrodynamical simulations cannot cover large-enough volumes while reaching small-enough halo masses~\cite{Kannan:2021xoz}. Semi-numerical simulations (such as {\tt 21cmFAST}~\cite{Mesinger:2010ne,Munoz:2021psm}), which rely on sub-grid models,  are instead commonly used. Detailed calibrations and comparisons between these different approaches are currently lacking, and will be critical to interpret upcoming data.

\item {\bf Consistent galaxy formation model.} Connecting galaxy properties to the underlying dark matter structure in a way that reproduces observed correlations between multi-wavelength observables is a major challenge. Sufficiently robust hydrodynamical simulations are capable of making predictions for such correlations, but are too expensive to run in large volumes. As such, development of galaxy formation models that can be applied on dark-matter-only simulations while accounting for the correlations between neutral and ionized gas, stars and dust in galaxies and galaxy clusters will be necessary.

Multi-probe simulations should also offer predictions for the intrinsic shapes of galaxies, another example of a very small-scale observable that is extremely difficult to model. The correlations of these shapes, known as their intrinsic alignments (IA), is an important systematic effect for next generation weak lensing surveys, but also contain information on galaxy formation and fundamental physics. Currently, galaxy shapes are either drawn from semi-empirical models, which require both high mass resolution simulations and extensive observations \cite{Joachimi2013}, or are obtained from hydrodynamical simulations \cite{Bate2020, Tenneti2021} which are computationally infeasible to be run with the required volumes. New techniques to rapidly assign realistic shapes to galaxies without incurring significant additional computational costs should be explored as an alternative, and outputs stored from future simulations should include the required quantities.

\item{\bf Neutrinos.} Neutrino oscillation measurements have shown that at least two of the three mass eigenstates of the Standard Model neutrinos are massive \cite{Zyla:2020zbs}. Massive neutrinos produce scale-dependent suppression of cosmic structures, with the largest effects on small scales, allowing for constraints on the total mass of neutrinos from cosmological measurements. Simulating massive neutrinos, which make up a non-negligible fraction of the total energy budget of the Universe, can be challenging since they decouple when relativistic, and have a free streaming scale of $\sim \mathcal O (1 h^{-1}{\rm Gpc})$. On smaller scales, their thermal velocity distribution needs to be accounted for in a structure formation calculation, unlike the CDM component. In an N-body approach, therefore, the six-dimensional distribution function of neutrinos needs to be sampled, i.e.~that at each location, an ensemble of neutrino particles should be initialized with the momentum distribution given by a Fermi-Dirac distribution.  This is a fully non-linear approach and represents a ``gold standard'' in the field, but suffers from Poisson noise unless the number of neutrino particles is prohibitively large \cite{Banerjee2018}.  Some recent approaches on how to reduce this noise include better sampling of neutrino momentum directions \cite{Banerjee2018}, hybrid fluid and N-body techniques \cite{2016JCAP...11..015B}, and by sampling only the deviations from the linear solution with particles \cite{Elbers2021}.  A computationally more efficient, albeit approximate method, is to model massive neutrinos with linear or perturbative approach which is then added to the large-scale non-linear gravitational potential in a simulation (see, e.g. Refs.~\cite{Brandbyge2009,2013MNRAS.428.3375A,Upadhye2016,Senatore2017}).  Depending on the mass of the individual neutrino species being simulated, this approximation eventually break down at sufficiently late times and sufficiently small scales since it lacks nonlinear evolution of neutrino perturbations as well as the back-reaction of non-linear matter on the neutrinos.  For small neutrino masses, which are usually of primary interest, these effects may not be significant \cite{Pedersen2021, Bayer2021}, but need to be calibrated carefully, depending on precision targets set by the sensitivity of future surveys.

\item {\bf Ray tracing.} With currently available ray tracing algorithms (see e.g.~Ref.~\cite{Hilbert2020}), it is computationally infeasible to cover both the large volume required by future weak lensing surveys, and yet maintain the accuracy at small scales required for strong lensing. Therefore, we must develop a multi-resolution ray tracing algorithm that will effectively cover the two regimes.

\item {\bf Baryonic effects in large-volume simulations.} Baryonic feedback effects are known to alter the local matter density and hence the weak lensing observables \cite{Schneider2019, Chung2020}. This is one of the leading systematic effects in cosmic shear analyses that is limiting the extraction of information from small scale measurements~\cite{Huang2021}. Modeling gas dynanics and feedback is also a crucial aspect of predicting the SZ signals, which depend on the ionized gas density/temperature at small scales \cite{Shaw2012,Park2018}. Therefore, these effects must be included in the modeling for future analyses. While attempts have already been made in existing hydrodynamical simulations, the predictions vary significantly due to lack of predictive control over the relevant astrophysical processes.
\end{itemize}

\section{New Physics Modeling Needs}
Cosmological simulations need to advance in terms of increased resolution, larger volumes, and better treatments of known physics, as described in the previous section. Additionally, as the observational reach of the surveys expands, they can be used to explore previously unconsidered physics regimes. It is therefore only natural that simulation methods be developed to model these new probes.

\begin {itemize}
\item {\bf Modified Gravity.}
The current theory of gravity is given by Einstein's theory of General Relativity (GR), and the currently leading explanation for the observed accelerated expansion of the universe is the cosmological constant, $\Lambda$, which is supported by all current observations. While the cosmological constant is mathematically simple idea, it is extremely unnatural from a theoretical physics standpoint \cite{Weinberg1989}.  Alternative models for accelerated expansion roughly split into two categories: dark energy and modified gravity. Dark energy models add an energy component with equation of state $w \neq -1$, and these are straightforward to simulate using virtually any existing code as this amounts to a modification of the background expansion rate.  The situation is quite different with modified gravity models. There are many proposed models (see for example Ref.~\cite{Clifton2012}), and they generically modify the Poisson equation in a non-trivial way by introducing non-linearity. This makes modified gravity simulations much more computationally expensive than \LCDM simulations, although algorithmic improvements and physical approximations used in recent modified gravity codes help reducing the cost \cite{Bose2017,Arnold2019,Ruan2021,Hernandez2022}.  Additional complication is that different models of modified gravity generally require different simulation setups.  Nevertheless, understanding theory of gravity remains a fundamental question in Physics, and testing gravity on cosmological scales (for a comprehensive review, see Ref.~\cite{Ishak2019}) will continue to be one of the primary science goals of the upcoming generation of large-scale structure observations \cite{Belgacem2019,Alam2021}.

\item {\bf Dark Matter Models.}
Most cosmological N-body and hydrodynamical simulations have focused on modeling particle dark matter that is cold, collisionless, and stable, as part of the $\Lambda$CDM paradigm. However, since the microphysical nature of dark matter, or even possibly a complicated dark sector, remains unclear, particle theorists have proposed a landscape of dark matter candidates with mass across tens of order of magnitudes~\cite{Battaglieri:2017aum}. Many of these candidates demand different simulation approaches from that of the cold dark matter (CDM) given their different properties. Below we list some examples of dark matter candidates and their related simulation demands or challenges, which are further summarized in a companion white paper focusing on cosmological simulations for dark matter physics \cite{DMsimsWP}.
\begin{itemize}
    \item {\bf Warm dark matter.} Warm dark matter (WDM) is a family of models with sizable thermal motions, in between that of CDM and (ruled out) hot dark matter at the first epoch of structure formation. It is associated with a free-streaming length that washes out small structures below the length, which leads to a cutoff in the matter power spectrum~\cite{Bode:2000gq}. Examples of WDM include sterile neutrinos and gravitinos from SUSY theories~\cite{Viel:2005qj,Drewes:2016upu}. WDM has been constrained from the Lyman-$\alpha$ forest~\cite{Irsic:2017ixq}, Milky Way subhalos, and the 21-cm signal~\cite{Schneider:2018xba}. But many of those studies suffer from systematic uncertainties related to baryons. For example, the constraint from Lyman-$\alpha$ forest data strongly depends on the modeling of the intergalactic medium, such as its temperature fluctuations~\cite{Hui:2016ltb}. Dedicated hydrodynamic simulations will be helpful to reduce the systematic uncertainties.
    
    \item  {\bf Interacting dark matter.} Interacting dark matter (IDM) candidates that strongly interact with Standard Model particles such as protons, neutrons, or electrons. For some part of the parameter space, the interaction is so strong such that IDM cannot be probed by direct-detection experiments due to the overburden from the Earth's atmosphere or crust. Therefore, cosmological observations are one of the most sensitive probes for IDM, including the CMB, the Lyman-$\alpha$ forest, and Milky way subhalos~\cite{Buen-Abad:2021mvc}.

    \item {\bf Self-interacting dark matter.} Self-interactions are ubiquitous for dark matter models, especially when dark matter is a part of the dark sector.  Sizable self-interactions of dark matter, with cross section strength at $\mathcal{O}(1\,\text{cm}^2/\text{g})$, may address the so-called ``small-scale problems" of $\Lambda$CDM while keeping its success on predicting the large-scale structure~\cite{Tulin:2017ara}. The self-interactions of the dark matter can be diverse, e.g., elastic, dissipative, velocity-dependent, forward interactions, but many numerical  studies only capture a small subset with phenomenological descriptions. The central region of self-interacting dark matter halos may experience dramatic changes in structure given the gravothermal collapse or dark matter-ordinary matter interactions. But this region is often omitted in numerical simulations because of the high computational cost.
    
    \item {\bf Dissipative dark matter.} If dark matter is connected to other light dark sector particles, its self-interactions can emit those particles and become dissipative.  Similar to ordinary matter, dissipative dark matter could experience cooling and heating through interactions with the environment. As a sub-component of the total dark matter, it could also fragment into dark clumps or form dark disks if the cooling effect is strong. Dedicated hydrodynamic simulations are needed to study dissipative dark matter. 
    
    \item {\bf Decaying dark matter.} Dark matter particles can be long-lived yet unstable. They could decay into Standard Model particles (e.g. sterile neutrinos) or other dark matter/dark sector particles on a long-time scale. High-resolution cosmological N-body simulations are often employed in studies of decaying dark matter~\cite{Wang:2014ina,Hubert:2021khy, Mau:2022sbf}.  Hydrodynamical simulations are also needed to study the impacts of processes such as  baryonic feedback~\cite{Wang:2014ina,Hubert:2021khy}. 
    
    \item {\bf Ultralight dark matter (fuzzy dark matter).}  Dark matter can be made of ultralight scalar, psudo-scalar, or vector particles. They collectively behave as a classical wave given their high occupation number~\cite{Hui:2021tkt}. Ultralight dark matter candidates are featured in many beyond-Standard-Model scenarios as the pseudo-Nambu-Goldstone-Boson of the broken symmetries. Examples include fuzzy dark matter~\cite{Hu:2000ke,Marsh:2015xka,Hui:2016ltb}, QCD axions, axion-like-particles, and dark photon dark matter. Ultralight dark matter suppress  small structures on the scale below the de Broglie wavelength. Thus it can be probed by observations such as Lyman-$\alpha$ forest, MW subhalos, or the formation of the first galaxies at cosmic dawn~\cite{Irsic:2017yje,Schutz:2020jox,Munoz:2019hjh,Jones:2021mrs}. 
    
    Numerical simulations for ultralight dark matter include: 1) Schr\"{o}dinger–Poisson equations that govern the evolution of the wave function of the ultralight dark matter ~\cite{Schive:2014hza,Schive:2014dra,Veltmaat:2018dfz}; 2) N-body simulations based on the Schr\"{o}dinger-Vlasov correspondence (for scales much greater than de Broglie wavelength)~\cite{Widrow:1993qq}; 3) fluid simulations based on the Madelung-transformed Schr\"{o}dinger–Poisson equations~\cite{Niemeyer:2019aqm}.  Few simulations of ultralight dark matter go beyond the dark matter only simulations to include baryons. Additionally, higher resolution simulations over wider ranges of parameters are needed.
    
    The cosmological evolution of QCD axion dark matter can be classified into two scenarios: (a) Peccei-Quinn symmetry is broken before or during cosmic inflation, and (b) broken after inflation. While the production process of axion dark matter of scenario (a) is relatively easy to model, that of scenario (b) requires numerical simulations given the production of the topological defects in the intermediate stage. Dedicated high-resolution numerical simulations have been recently developed to accurately track axion dark matter abundance~\cite{Gorghetto:2018myk,Eggemeier:2019khm,Gorghetto:2020qws,Buschmann:2021sdq} for scenario (b). 

    \item {\bf Ultraheavy dark matter.}  Dark matter can be made of ultraheavy objects
    with mass from around the Planck mass to solar masses. Examples of ultraheavy dark matter include primordial black holes, massive compact halo objects, exotic compact objects~\cite{Giudice:2016zpa}, and dark matter blobs~\cite{Diamond:2021dth}. In the absence of strong self-interactions or interactions with ordinary matter, probes of ultraheavy dark matter are limited to gravitational probes such as micro-lensing or graviational waves produced from merging binaries. These probes can be strongly affected by the distribution of ultraheavy dark matter~\cite{Carr:2021bzv, Giudice:2016zpa,Diamond:2021dth}, motivating dedicated numerical studies of the clustering of ultraheavy dark matter. 
    
    \item {\bf Multiple dark matter components.} Given the landscape of the dark matter candidates, it is easy to imagine that the dark matter consists of multiple components. For example, CDM can be the dominant component, while other dark matter candidates are sub-dominant. Examples of this scenario include axiverse~\cite{Arvanitaki:2009fg} and cannibal dark matter~\cite{Carlson:1992fn}. An important question is the distribution of the sub-dominant component inside dark matter halos. Just as for the distribution of baryonic matter and dark matter, the distribution of the sub-dominant component inside the halo may not be a simple re-scaling of the dominant component. Dedicated numerical simulations are needed to pin down the distribution of the sub-dominant components and make the predicted signatures reliable (e.g.~\cite{Anderhalden:2012qt,Banerjee:2022era}).
    
\end{itemize}
\end{itemize}

There are other new ideas on dark matter,  motivated by modified gravity theories, such as superfluid dark matter~\cite{Khoury:2021tvy} and the apparent dark matter from entropic gravity~\cite{Verlinde:2016toy}. Many of those models still lack dedicated cosmological simulations or simulations with baryons.

\section{Statistical Inference and Simulation Suites}

The current cosmological Standard Model, $\Lambda$CDM, is an excellent fit to the data but has several theoretical shortcomings and is generally perceived, very like the particle physics Standard Model, to possess only a transitory existence, and be eventually replaced by a more complete description. But because $\Lambda$CDM is so successful, deviations from it will be subtle and difficult to nail down. Consequently, the next generation of cosmology experiments will be driven not only by the accumulation of statistics but also by the need to understand, mitigate, and control systematic uncertainties. To make substantial headway in the latter task, the ability to create detailed and realistic ``virtual universes" on demand is gaining central importance, so much so, that the ultimate scientific success of upcoming sky surveys hinges critically on the success of the modeling and simulation effort.

Scientific inference with sky surveys is a statistical inverse problem, where, given a set of measurement results, one attempts to fit a class of physical models to the data (which include models for the observational process), and to infer the values of the model parameters. Typically, such analyses require many evaluations over a very large number of ``virtual universes". The main difficulty lies in the fact that producing each virtual universe requires, in principle, an extremely expensive numerical simulation carried out at high fidelity. Emulators are effectively fast surrogate models that can be used as an alternative route to solving this inverse problem \cite{Heitmann2006}.

\subsection{Emulating the Observable Universe}

The importance of providing predictions for cosmological surveys via emulators is now widely recognized; emulation-based predictions have become very popular over the last few years~\cite{Lawrence2010,Heitmann2014,Lawrence2017,Bocquet2020,McClintock2019,Zhai2019,McClintock2019b,Nishimichi2019,Kobayashi2020,Mootoovaloo2020,Heydenreich2021,Knabenhans2021,Takhtaganov2021}. It is now possible to carry out high-quality simulation suites that provide the input to the emulators for a range of cosmological statistics, such as halo mass functions, matter power spectra, and halo bias. (Some of the emulators have gone beyond $\Lambda$CDM as well.) However, in order to fully integrate the emulators into the analysis frameworks used by the surveys to extract cosmological parameters, it is very desirable to create emulators connected directly to the survey observables. As a concrete example, the cluster mass function is commonly used to derive cosmological constraints, but it is not a quantity that can be easily extracted from the observations. All measurements, including weak lensing shear, do not directly provide mass measurements, but rather approximations or proxies for an idealized ``cluster mass". The translation from the observable to the mass function from simulations adds additional uncertainties into derived cosmological parameters and could be avoided if emulators would directly predict the quantity of interest, which is the cluster abundance, measured in a way that is most relevant to how the survey is actually carried out.
This level of forward modeling would involve new simulation and analysis efforts and the development of more flexible and sophisticated emulation approaches. Given that a successful implementation can potentially eliminate a major source of uncertainty and bias, this is clearly a worthwhile step. With a broad enough simulation footprint, such an approach would also enable easier connections across measurements carried out in different wavebands. 

\subsection{Extending the Physics Content of Simulations}

Most emulator efforts so far have relied upon gravity-only simulations. These are an order of magnitude less expensive than hydrodynamic simulations, and yet carrying out a high-quality suite of gravity-only simulations has only become possible in the last few years due to increases in available computing resources. For hydrodynamics simulations, such campaigns are still out of reach because of the much larger time to solution per simulation. Additionally, 
hydrodynamics simulations have many modeling (nuisance) parameters and uncertainties, increasing the design space for the emulation and in turn increasing the number of simulations to be carried out. (For gravity-only simulations, after having established criteria for precision simulations, only the fundamental cosmological parameters have to be varied, keeping the simulation campaign size manageable.)  With the advent of exascale supercomputing resources (and beyond) in the coming years, and employing strategies such as multi-fidelity simulations, this problem can be significantly reduced, assuming the hydrodynamics codes can take full advantage of the new generation of architectures. If this turns out to be the case, many opportunities open up: Emulators can be built to investigate and optimize subgrid model parameters, be deployed to gain a better understanding of the interplay of subgrid model and cosmology parameters, and to directly predict observational quantities for different surveys accessing different wavelength regimes.

\subsection{Robust Error Estimation and Parameter Exploration}

Error estimation with emulators is a potentially very powerful avenue of research but remains to be properly realized in many of the current generation of emulators. Partly this is because error estimation is inherently difficult and partly because the methods used have been too informal, and insufficiently sharp. For example, there is no rigorous theory for error convergence and systematically handling discrepancy between model predictions and observational measurements remains an open problem. Because the formal statistical uncertainties in the observations are reducing with time, the onus is on modeling systematic errors, including the errors in the emulators. This area is relatively little-studied in the statistics literature although there are some useful investigations in discrepancy modeling~\cite{Brynjarsdottir2014}; the power of the results obtained, however, is relatively limited.

Another problem is that the dynamic range in cosmology is vast and it is computationally impractical to model all the relevant processes via a first principles approach. Consequently, some of the inputs in the subgrid models must be empirical, based on known results from observations. This adds another layer of complexity to error estimation because the proper treatment of such evidence in a cosmological analysis potentially requires a separate set of investigations for each empirical input. However, we note that continuous inclusion of observational data in emulator construction will be helpful in reducing the volume of parameter space that needs to be explored. (As mentioned previously, multi-fidelity simulations are also useful here in minimizing the amount of computational work.) Adaptive sampling methods are very useful in time-domain applications and they can be easily transplanted to cosmology, provided error analyses can continue to be undertaken in a robust manner.

\section{Future of High Performance Computing}
\vspace{-0.3cm} 

\subsection{Next-generation Supercomputing Platforms}
\vspace{-0.2cm} 
The arrival of the first generation of exascale supercomputers, Aurora and Frontier, at the Leadership Computing Facilities at Argonne and Oak Ridge National Laboratories provides an extraordinary opportunity to push scientific simulations to the next level. In cosmology, they enable two classes of simulations relevant to cosmological surveys: gravity-only simulations with unprecedented volume coverage and resolution and hydrodynamics simulations with exceptionally detailed and realistic modeling of baryonic physics in the Universe. The Exascale Computing Project (ECP) led by the Office of Advanced Scientific Computing Research (ASCR) in collaboration with other DOE science program offices has made tremendous strides to prepare scientific applications for these resources and will continue to do so~\cite{ecp}. As part of this effort, important challenges have been identified, including the efficient use of computational accelerators, effective memory access patterns, performance portable programming models and scalable algorithms. For gravity-only simulations, some of these challenges have already been successfully addressed by a subset of codes~\cite{Habib2016, Potter2017, Garrison2021}.  For hydrodynamics simulations, these challenges are far more complex, but are being tackled by different codes \cite[for example, see][]{Sexton2021, Springel2021, Frontiere2022}. Continuous developments on these fronts are extremely important in order to enable full use of exascale systems and the ones that will follow them.

\subsection{Scalable Analysis Approaches}
\vspace{-0.2cm} 
Scalable analysis approaches are as important as the development of the simulation codes. In principle, many petabytes of data can be easily generated by current and next-generation HPC systems, in practice, however, storage capacities are limited and the handling and processing of very large data sets require large supercomputing resources in their own right. Consequently, carefully designed analysis routines have to be instantiated on-the-fly while the simulation codes themselves are running (``in situ" analysis). The development of these analysis routines faces the same challenges as the simulation codes, and scalability and efficient usage of the available architectures are mandatory \cite{foresight2020, foresightGPU}. A successful cosmological simulation program therefore needs to ensure that the development of the codes and the analysis routines go hand in hand. In particular, these analysis routines can leverage a tight coupling with simulation codes to best balance available memory and compute capabilities of the supercomputing system. This co-development of simulation codes and analysis routines is complicated by the fact that cosmological simulations aim to provide predictions for a wide range of observations. A carefully orchestrated analysis approach has to be developed with cosmological surveys in mind -- close collaboration between simulators and observers is essential for its success.

\subsection{Verification and Validation}
\vspace{-0.2cm} 
The accuracy requirements for cosmological simulations are stringent. As outlined above, the simulations provide the foundation for the analysis of current and next-generation cosmological surveys. Given the aim to constrain, e.g., dark energy parameters at the percent level, simulations and the coupled analysis and modeling approaches have to deliver results at least at the same level of accuracy, and better, if possible. The community has made good progress with regard to code verification in the last few years by carrying out rigorous comparison projects \cite{Heitmann2005,Schneider2016,Onions2012,Agertz2007} and convergence studies \cite{Heitmann2008}. However, not all differences between the codes and analysis tools have been fully resolved and/or understood. In particular, in the area of hydrodynamic simulations, much more work is needed to obtain the desired levels of robustness, although there is recent evidence of progress in this direction~\cite{nifty,frontiere}.

Validation (confirming the accuracy of the simulation predictions by direct comparison against observations) is another crucial area that requires a concerted effort between different code and analysis development teams and observers. The upcoming surveys will provide a rich data set for this effort. A delicate issue is how to control errors coming from empirical modeling used within the setup of the simulations. The detailed connection between the simulations and the survey observables has to be tightened up considerably as this is the most problematic aspect of the validation program from the simulation perspective.

\section{Conclusion}

Cosmological surveys carried out over the next decade are poised to make discoveries that will either extend or confirm the $\Lambda$CDM model. Both alternatives are significant -- in the first instance, observational input in finding ``Beyond $\Lambda$CDM'' corrections is clearly of fundamental importance, and in the second instance there will be a sharp reduction in the number of possible alternatives to the model, with ramifications for future tests and other investigations. To achieve the level of accuracy that is desired, close coupling to a state-of-the-art simulation campaign that not only provides a complete and robust modeling platform for each survey but also provides a capability to simultaneously model a number of observations from a range of facilities, including their cross-correlations, is required. Next-generation HPC systems promise to provide a capability that can help achieve these goals; getting to the desired results will require a concerted effort in implementing new algorithms/models, and evolving the simulation codes and associated analysis tools. Additionally, close collaborations with survey teams will be an essential element for success.

Over the course of the next decade, we expect exciting discoveries to be made by combining and analysing data sets from large-scale structure, CMB, and line intensity mapping experiments. In preparation, we must develop simulations with a broad set of observables, including their correlations, in order to conduct these analyses. However, despite their importance, resources to aid in developing these simulations and associated analysis methods have been scarce since 1) they do not belong to a specific collaboration/telescope, and 2) to generate fully coherent simulations, expertise from various disparate areas is required. This situation will have to evolve in a positive direction, if we are to achieve the full scientific potential of future surveys.

\bibliographystyle{unsrt}
\bibliography{main.bib}

\end{document}